\documentclass[twocolumn,showpacs,preprintnumbers,aps]{revtex4}

\usepackage{graphicx}
\usepackage{dcolumn}
\usepackage{bm}

\begin{document}

\title{DNA hybridization to mismatched templates: a chip study}

\author{Felix Naef$^{1}$, Daniel A. Lim$^{2}$, Nila Patil$^{3}$, and Marcelo Magnasco$^{1}$}

\affiliation{$^{1}$Center for Studies in Physics and Biology, Rockefeller University, 1230 York Avenue, NY 10021}
\affiliation{$^{2}$Laboratory of Neurogenesis, Rockefeller University, NY.}
\affiliation{$^{3}$Perlegen Inc., 3380 Central Expressway, Santa Clara, CA 95051.}

\date{\today}

\begin{abstract}
High-density oligonucleotide arrays are among the most rapidly expanding
technologies in biology today.
In the {\sl GeneChip} system, the reconstruction of the target concentration
depends upon the differential signal generated from hybridizing the target
RNA to two nearly identical templates: a perfect match (PM)
and a single mismatch (MM) probe.
It has been observed that a large fraction of MM probes
repeatably bind targets better than the PMs,
against the usual expectation from sequence-specific hybridization; 
this is difficult to interpret in terms of the underlying physics.
We examine this problem via a statistical analysis of a large set 
of microarray experiments.
We classify the probes according to their signal to noise ($S/N$) ratio, defined as the 
eccentricity of a (PM, MM) pair's `trajectory' across many experiments.
Of those probes having large $S/N$ ($>3$) only a fraction behave
consistently with the commonly assumed hybridization model.
Our results imply that the physics of DNA hybridization in microarrays
is more complex than expected, and they suggest new ways of constructing 
estimators for the target RNA concentration.
\end{abstract}

\pacs{87.15.-v, 82.39.-k, 82.39.Pj}

\maketitle

Interest in the detailed physics of DNA hybridization is rooted in
both purely theoretical and practical reasons.
Studies of the denaturing transition started with models of perfectly
homogeneous DNA \cite{PB1}, soon followed by studies of
sequence-specific disorder \cite{HWA,ZHANG,NELSON}.
The specificity with which DNA binds to its exact complement
as opposed to a mismatched copy (a ``defect'') has been studied 
experimentally \cite{DEFECT,BONNET} and theoretically
\cite{SALERNO,SINGH,WATTIS}. In this context it has been found that a 
fair fraction of the energetics of DNA hybridization is related to 
{\em stacking} interactions between first-neighbor bases, in addition to 
the obvious strand-strand contact \cite{STACK,STACK2}. In this paper we
present a study of mismatch hybridization stemming from a very practical
problem, hybridization in DNA microarrays. We shall show experimental 
evidence that the system displays behavior which appears to be hard to
account for on the basis of the extant view of hybridization specificities. 

DNA microarrays provide an experimental technique for measuring thousands
of individual mRNA concentrations present in a given target mixture. They are
made by depositing DNA oligonucleotide sequences (probes) at specific
locations on solid substrates. The probes can be either pre-made sequences as
in cDNA spotted arrays, or they can be grown \emph {in situ}, letter by letter,
as in high-density oligonucleotide arrays \cite{GC1}.
The target mRNA is amplified (into either cDNA or cRNA depending on the protocol)
and the product labeled fluorecently before being hybridized onto the array. The spatial
distribution of fluorescence is then measured using a laser, providing estimates
for the target concentrations. In {\sl GeneChip} arrays, the synthesis of probe
sequences by photolithographic techniques requires
a number of different masks and deposition processes per added base,
making it impractical to grow more than a few dozen 
nucleotides. For such lengths, hybridization specificity is not expected to
be high enough.
To solve this conundrum, {\sl GeneChip} technology is based on a two-fold approach,
involving {\em redundancy} and {\em differential signal} \cite{GC1,GC2,GC3}.
First, several different sequence snippets (each 25 bases
long) are used to probe a single transcript; and second, each of these probes
comes in two flavors. The perfect match (PM) is perfectly complementary to a
portion of the target sequence whereas the single mismatch (MM) carries a
substitution
to the complementary base at its middle (13th) position. The rationale behind
MM sequences is that they are expected to probe for non-specific hybridization
in a manner that we shall detail below.

In current incarnations of the chips,
each gene is probed by 14-20 (PM,MM) pairs (a probeset),
and the task is therefore to reconstruct a single number
(the target concentration) from these 28-40 measurements.
There are conceivably many ways in which this can be done, with various degrees
of noise rejection. The standard algorithm provided in the software
suite \cite{AFFYSOFT} offers one method.
However, as independent experimental techniques for measuring mRNA concentrations
(like northern blots) provided clues that the analysis process
should be improved upon, many researchers attempted to do so;
it was then discovered that a fair number of MM probes consistently
report higher fluorescence signal than their PM counterpart \cite{US}.
This observation is most intriguing because it violates
the standard hybridization model as outlined below. 
Thus, the notion that the specific binding signal alone can be obtained
as a differential of the PM and MM signals appears to fail in a subset of the
probes.

We shall show below that it is not a matter of a few stray probes,
by carefully examining the statistics of PM-MM pairs.
These statistics show that \emph{most}
of the probes misbehave to various degrees. Only a fraction of them, having
$MM>PM$, exhibit a flagrant violation of the basic assumptions, however,
these just point most obviously at the symptoms of a deeper problem which affects
all probes. Given the number of laboratories who are currently carrying out
such hybridization experiments, squeezing out even a meager extra bit of signal
to noise ratio from the data would be very valuable. It has
become clear that this shall not happen in the absence of a better understanding
of DNA hybridization to slightly mismatched templates. We shall
now attempt the first step toward this goal, which is to characterize
the problem. 

\begin{table}[b]
\caption{
Statistics of probe pairs with $MM>PM$ taken across a large
{\sl GeneChip} data collection. ``\%PS with $>1$'' means
``percent of probesets with
more than one $ MM>PM $ pair''. The yeast chip (last column)
is \emph{noticeably} different and better behaved than the other cases.
}
\vspace{.2cm}
\begin{tabular}{c|ccccc}
\hline\hline
Chip&
Dros&
HG-U95A&
Mu11K&
U74A&
YG\_S98\\
\hline  
{\# pairs per PS}&
14& 16& 20& 16& 16\\
chips analyzed&
36& 86& 24& 12& 4\\
{\% $ MM>PM $ }&
35& 31& 34& 34& 17\\
{\% PS with $>1$}&
95& 91& 95& 92& 73\\
{\% PS with $>5$ }&
58& 56& 71& 64& 21\\
{\% PS with $>10$ }&
4& 7& 26& 10& 2\\
\hline\hline 
\end{tabular}
\end{table}

\begin{figure}[htbp]
\resizebox*{.49\columnwidth}{!}{\includegraphics{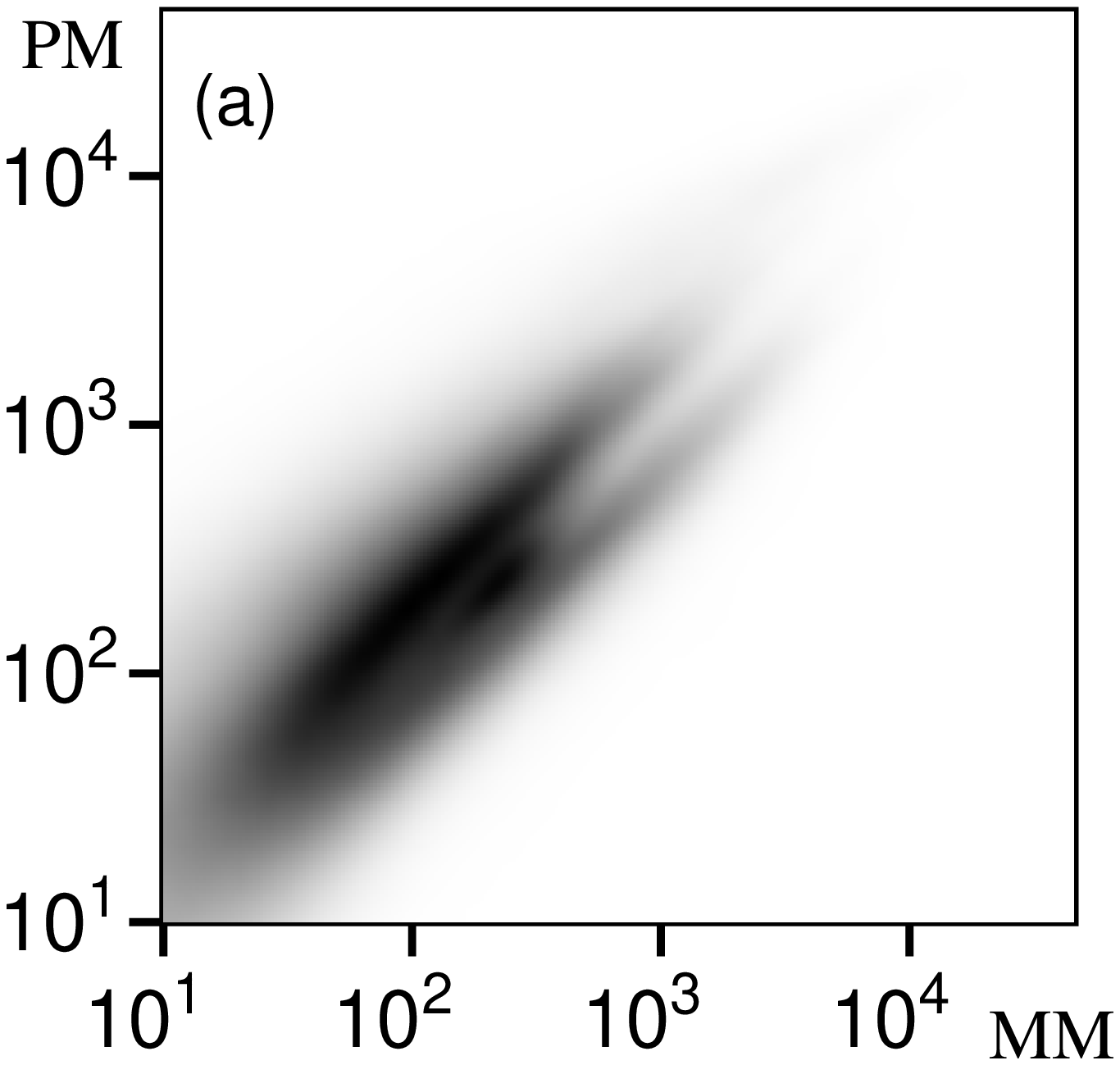}}
\resizebox*{.49\columnwidth}{!}{\includegraphics{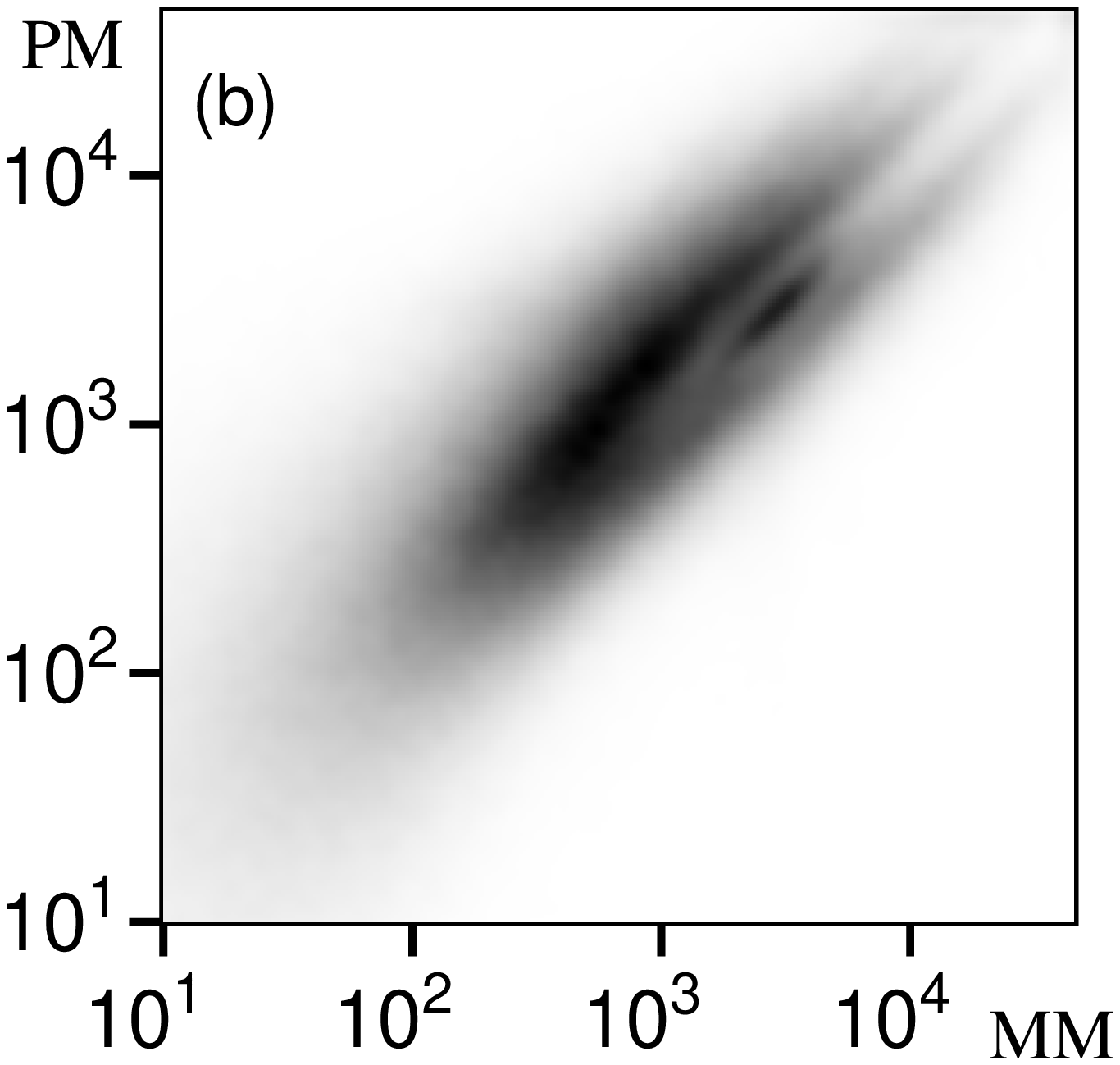}} 
\caption{
Joint probability distribution $ P(\log PM,\log MM) $
for two large datasets after background subtraction. (a) 86 HG-U95A human chips,
human blood extracts. (b) 24 Mu11K/A mouse chips, mouse brain extracts. Please
notice that three obvious features are present in both: the probability cloud
forks into two lobes at high intensity, and an intense ``button'' lies between
the two forks right in the middle of the range. Notice that the lower lobe
is {\em completely} contained below the diagonal $MM=PM$. 
}
\end{figure}

The rationale behind the use of MM probes is contained in the standard hybridization
model \cite{WONG}:
\begin{eqnarray*}
PM & = & I_{S}+I_{NS}+B\\
MM & = & (1-\alpha )\, I_{S}+I_{NS}+B\\
PM-MM & = & \alpha I_{S}
\end{eqnarray*}
Here $PM$ ($MM$) are the measured brightness of the PM (MM) probe,
$I_{S}$ the contribution from specific complementary binding, $I_{NS}$
the amount from nonspecific binding assumed to be insensitive to the substitution,
and $B$ a background of physical origin, i.e. the photodetector dark current
or light reflections from the scanning process. Then $\alpha $ is the reduction
of specific binding due to the single mismatch. These brightnesses are related
to the quantity of interest (the RNA concentration in the sample) through:
\begin{eqnarray*}
I_{S} & = & k\, [S]\\
I_{NS} & = & h\, [NS]
\end{eqnarray*}
where $[S]$ denotes the concentration of target RNA, $[NS]$ the
concentration
of whatever mixture contributes to the nonspecific hybridization. $k$ and
$h$ are probe dependent specific and nonspecific susceptibilities (possibly concentration
dependent) and include effects such as the areal density of probe,
various affinities, transcript length
dependent effects (longer transcripts are likely to carry more fluorophors depending
on the labeling technique). 

While it is no secret that the physics of hybridization is way more complex
than this simplistic model, one could still hope that it would essentially provide
a correct picture of {\sl GeneChip} hybridizations.
To summarize, let us outline the basic assumptions made so far: (i) non-specific
binding is identical in PM and MM, meaning that $ I_{NS} $ does not see the letter change;
(ii) $\alpha >0 $; (iii) $k$ and $h$ identical for PM and MM;
(iv) $k$, $h$ and $\alpha$ are reasonably uniform numbers across
a probe set. Notice that (i)+(ii) imply that $PM>MM$ always (see below). 

If $ PM-MM $ is not used as such, the background $ B $ needs to be subtracted
from the intensities, which can be done in a statistically
proper way as described in \cite{US}. 

According to the basic tenets of the standard model, it follows that $PM>MM$
for all probe pairs if the target RNA extract contains no sequences matching
exactly the MM. In reality, one observes a vast number of probe pairs for which
this assumption is violated; this behavior repeats consistently for a broad
range of conditions. Our experience is that most people in the know think of 
this problem in terms of an imperfect adherence to the standard model, or a bothersome
deviation from an otherwise properly behaving norm.
In other words, the way this problem is usually characterized
is "there's a number of probe pairs that don't work and we don't understand
why''. We shall show now that this is not so: the $MM>PM$ pairs are so
abundant that we like to propose the alternate view that
the model is simply inadequate for describing what actually happens, and that we do not
understand the basic physics of MM hybridization. Table 1 summarizes the statistics
for various chip series.

\begin{figure}
{\centering \resizebox*{0.49\columnwidth}{!}{\includegraphics{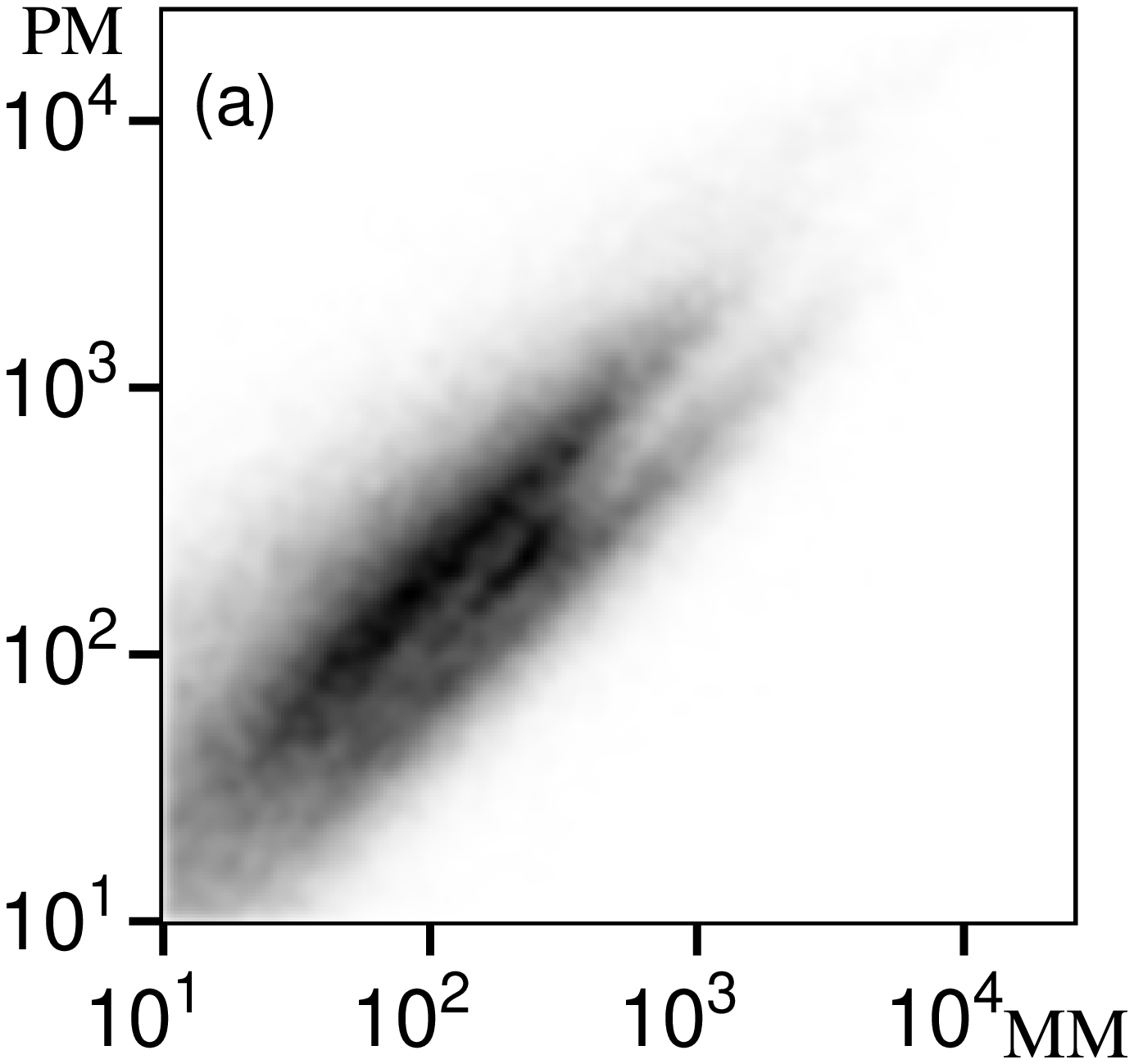}} 
\resizebox*{0.49\columnwidth}{!}{\includegraphics{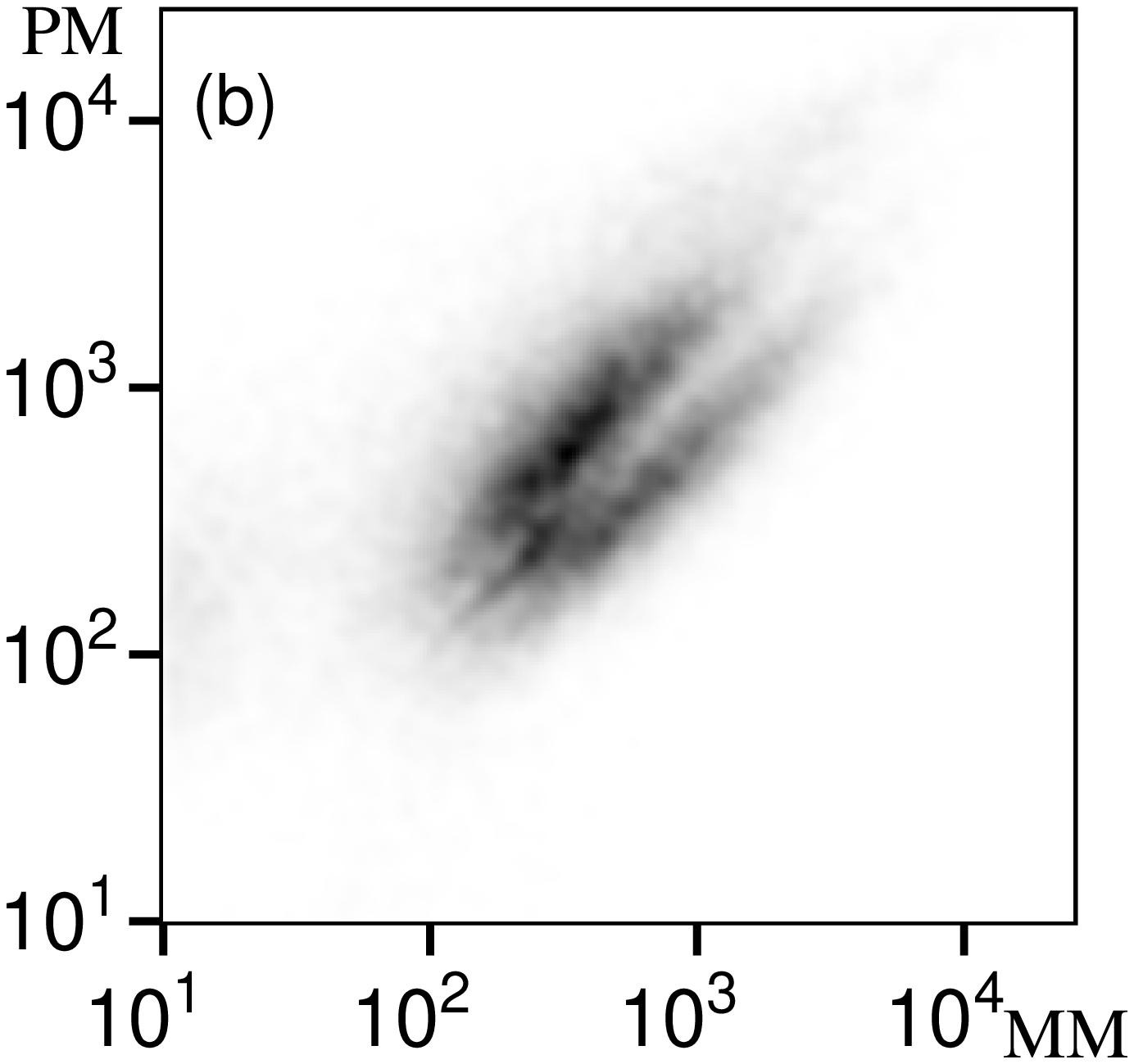}} \par}
{\centering \resizebox*{0.49\columnwidth}{!}{\includegraphics{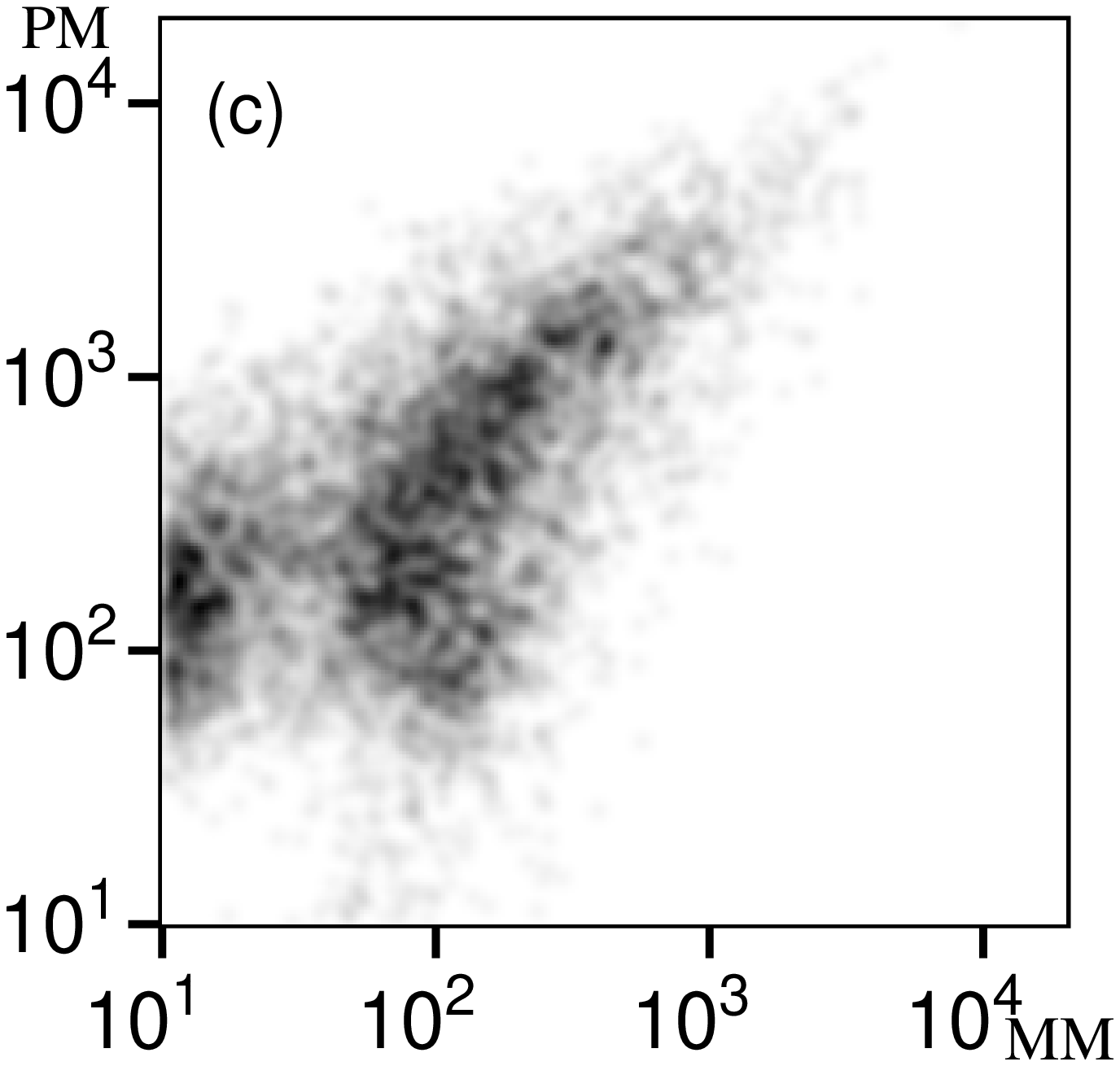}} 
\resizebox*{0.49\columnwidth}{!}{\includegraphics{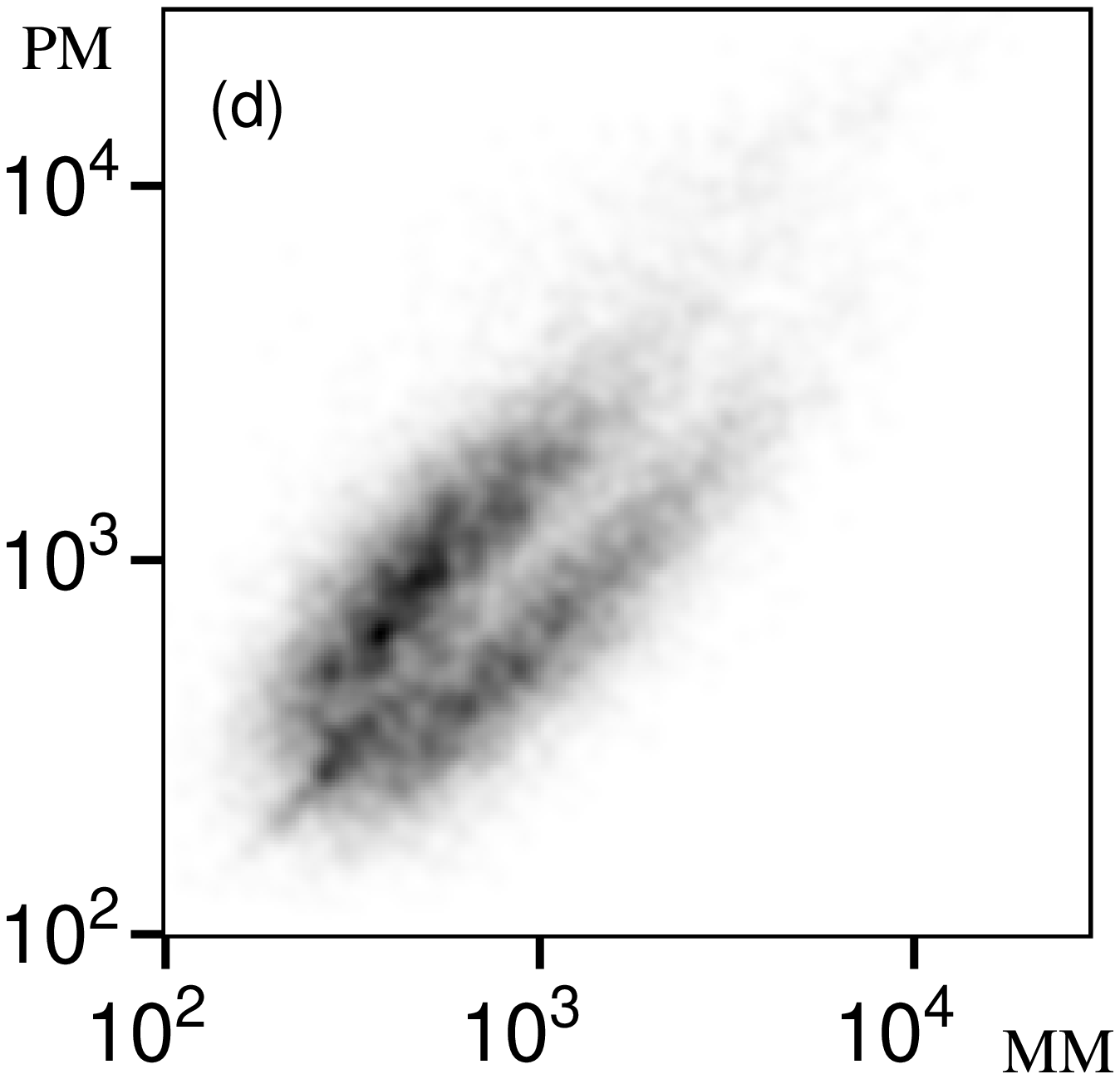}} \par}
\caption{
Histogram of probe center of mass. (a) All probes (to be compared with
Fig 1a). (b) Only those probesets with eccentricities $ e>3 $. (c) The probesets
of (b), further restricted to Large excursions ($ \lambda _{1}>0.168 $, the
top third of all probesets). (d) same as (c) for small excursions ($ \lambda _{1}<0.121 $,
the bottom third). Notice that (c) consists of all probe pairs with small S/N
and large signal, while (d) consists of pairs which both have small S/N
and small signal (bottom third). 
}
\end{figure}

The human HG-U95A chip series, for instance, has \~{} 400K probes for 12K different
probesets. Across a wide variety of conditions, we have observed approximately
$30\%$ of all probe pairs have $MM>PM$. This figure, by itself enormous, would be
easy to dismiss if most of them were in the low intensity range, where noise
is expected to be relatively higher and could conceivably be bigger than $|PM-MM|$,
or if they were clustered in a small set of problematic probesets. Neither is
true: $91\%$ of all probesets have at least 1 probe pair with $MM>PM$, and still
$60\%$ of probesets have 5 such probe pairs out of 16. In addition, the $MM>PM$
pairs are fairly distributed with respect to brightness (cf. Fig. 1). 

What could conceivably be the source for observing $MM>PM$? A perplexing extra
bit of information lies in a simple statistic,
the joint probability distribution
$ P(\log PM,\log MM) $.
According to the standard model, 
\[
\frac{PM}{MM}=\frac{I_{S}+I_{NS}+B}{(1-\alpha )I_{S}+I_{NS}+B}\]
If $ I_{s} $ dominates over $ I_{NS}+B $ then $ PM/MM\to 1/(1-\alpha ) $,
while if $ I_{s} $ vanishes (as when the transcript is just not present in
the sample) then $ PM/MM\to 1 $.

Thus we expect 
\[
1\leq \frac{PM}{MM}\leq \frac{1}{1-\alpha }\]
Thus, the standard model predicts that $ P(\log PM,\log MM) $ should be supported
in a band, with lower limit corresponding to the diagonal $PM=MM$ when
cross-hybridization dominates, and with an upper limit given by $MM=(1-\alpha )PM$
for fully specific binding. Naively one would further assume that for low brightness
most of the signal comes from nonspecific binding, while most would come from
specific binding for high brightness. Fig. 1 shows something quite otherwise:
as brightness increases, the joint probability distribution forks into two branches.
The crest of the lower one lies fully below the $MM=PM$ diagonal.

The characteristic shapes of $P(\log PM,\log MM)$ are likely signatures
of sequence-dependent effects. However, any hypothesis is impossible to verify
as the probe sequences are not released to the public. Nevertheless, there are
some obvious suspects. First, the nontrivial susceptibilities $k$ and $h$
mentioned above depend on the areal density of probe, which is sequence-dependent
by virtue of the varying efficiencies of the lithography process. Secondly,
nucleic acids need to unstack the single-stranded probes in order to form each
new duplex as they hybridize. Further, stacking energies are extremely sensitive
to sequence details, which might result in large energy barriers. This would
translate into kinetics constants that vary exponentially (\`a la Arrhenius) in
these energies, and lead to important consequences as the hybridization reactions
are not carried to full thermodynamic equilibrium.

Given a set of $N$ experiments, further insight can be obtained by following
a pair $\vec{P}_i=(\log PM_i,\log MM_i)$ with $i=1,\dots,N$
\emph{}\textit{\emph{across}} the entire dataset (after subtracting B). Ideally,
these points would fall on a curve parametrizable by the mRNA concentration.
In reality, however, the observed patterns range from nearly one-dimensional
to almost circular clouds. To classify probes, we computed the center of mass CM
and inertia tensor ${\cal I}$ of the set of points $\{\vec{P}_i\}$. The positive
eigenvalues of ${\cal I}$, $I_1\geq I_2$ define the eccentricity
$e=\sqrt{I_1/I_2}$ and
largest excursion $\lambda_1=\sqrt{I_1}$. Pairs with high eccentricities
are those carrying high S/N, whereas $e\sim 1$ characterizes
a very noisy probe pair.

Fig. 2 illustrate the distribution of center of mass
after different filtering for $e$ and $\lambda_1$.
It turns out that Fig. 2a looks very similar to Fig. 1, which is not a priori
evident. On the contrary, this similarity emphasizes that most probes behave
in a very reproducible manner. For instance, probes lying below the
$PM=MM$ diagonal at the high-intensity end do so in essentially all
of the 86 experiments (leading to a CM that is also below the diagonal),
instead of visiting different regions of the plot.
Another striking result is that (i) selecting for $e>3$ eliminates most of the low-intensity
probes (Fig. 2b), (ii) the remaining set contains two components: one consisting
of the large $\lambda_1$ probes (Fig. 2c) lying mostly in the $PM>MM$ region;
and the small $\lambda_2$ component forming an almost perfectly symmetric
``tulip" structure (Fig. 2d), containing two forked branches plus the button described
in Fig. 1.

Another troubling effect which deeply affects attempts at analysis is the very broad
brightness distributions within probes belonging to the same gene. Fig.
4 shows that the $PM$ probe intensities span up to four decades. Possible
reasons for such behavior are again sequence specific effect similar to those
discussed in the context of the MM behavior.

\begin{figure}
\resizebox*{1\columnwidth}{!}{\includegraphics{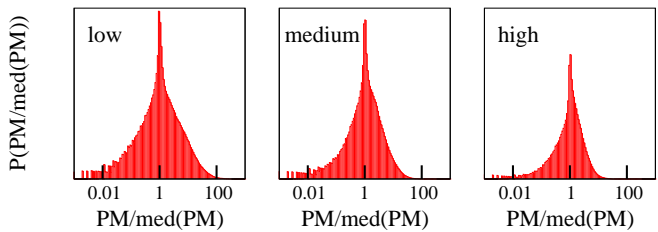}} 
\caption{
Relative $ PM $ intensity distributions within probesets
(after subtracting $ B $). The data shows the 86 HG-U95A human chips used previously.
Probesets are split into three groups according to their median PM intensity.
In all cases, the distributions of $ PM/\mbox {median}(PM) $ span up to four
decades. Notice there are signs of saturation in the right tail of the high-intensity
set.
}
\end{figure}

The main practical challenge is reconstructing the target mRNA concentration
from the probeset data. As we showed, the variability in the hybridization properties
of the probes is larger than naively anticipated, therefore, it is unlikely
that a single definitive procedure will be appropriate in all cases. On the contrary,
it is desirable to have several analysis tools at hand for viewing the data
from different angles. For instance, as a consequence of the strongly probe
dependent susceptibilities $\alpha,k$ and $h$, the differential $PM-MM$
will not consistently be a good estimator of the true signal. Given the unclear
information contained in the $MM$, one alternative we studied is not considering
them at all. The mRNA expression level is then obtained from a robust geometric
average of the PM-B values, after a careful estimation of B \cite{US}. The use
of geometric averages (rather that arithmetic) is dictated by the distributions
in Fig. 4. Of course, using only PM probes neglects cross-hybridization effects
that would be detectable by a working MM probe, and hence tends to be less sensitive
at the low-intensity end. One the other hand, it allows to rescue probesets with
a high number of misbehaving MMs.

A completely different approach, closer in spirit to the model-based method
\cite{WONG}, would be to extend the ellipsoid of inertia idea to the the full probeset.
Concretely, one would take the matrix
$A^{j}_{i}=(\log PM^{j}_{i},\,\log MM^{j}_{i})$
($j=1,\dots,N_p$ is the probe and $i$ the experiment index) and do a principal
component analysis to identify the modes carrying the most signal. After singular
value decomposition
$\hat{A}=U\, \Lambda \, V^{T}$, where $\hat{A}^{j}_{i}=A^{j}_{i}-m^{j}$
and $m^{j}=\frac{1}{N}\sum _{i}A^{j}_{i}$ is the center of mass, the signal
$s_{i}=\sum_j(m_{j}+\hat{A}^{j}_{i})\, V_{j}^{1}$ is given by the projection
onto the largest direction of variation. A signal-to-noise measure for the entire
probeset is given by
$\frac{S}{N}=\frac{\lambda_1}{\sqrt{\sum ^{N_p}_{j=2}\,\lambda^2_j}}$. Preliminary
testing of the method has lead to very promising results.

In conclusion, we showed that the hybridization of short length DNA sequences
to single mismatched templates exhibits a far more diverse picture than what is
usually assumed. These observations do not only point at interesting
physics in the DNA hybridization process to short sequences with defects, attached
to a glass surface; they also have strong consequences for designers of {\sl GeneChip}
analysis tools, especially when it comes to the level of noise rejection of
different methods. We hope this will bolster interest in the physics of
hybridization and mismatch characterization.

We'd like to thank E. van Nimwegen, E. Siggia, and S. Bekiranov for sharing interesting ideas.
MM acknowledges support of the Meyer Foundation; FN is a Bristol-Myers Squibb Fellow in Basic Neurosciences
and acknowledges support from the Swiss National Science Foundation.

\end{document}